\documentstyle [12pt,aaspp4,epsf]{article}
\def \ie {{\em i.e.,~~}}
\def\beq{\begin{equation}}
\def\eeq{\end{equation}}
\def\ber{\begin{eqnarray}}
\def\eer{\end{eqnarray}}

\def \tablerule {\noalign {\hrule}} 

\begin{document}

\lefthead {Bharadwaj, Sahni, Sathyaprakash, Shandarin and Yess}
\righthead {LCRS}

\title {Evidence for Filamentarity in the Las Campanas Redshift
Survey}
\author {Somnath Bharadwaj$^{\rm a}$,
Varun Sahni$^{\rm b}$, B.S. Sathyaprakash$^{\rm c}$, 
Sergei F. Shandarin$^{\rm d}$, Capp Yess$^{\rm e}$}
\affil {$^{\rm a}$ Department of Physics and
Meteorology,  and,  Center for Theoretical Studies,  
{I. I. T. Kharagpur 721 302, India }}
\affil {$^{\rm b}$Inter-University Centre for Astronomy \& Astrophysics,
Post Bag 4, Pune 411007, India}
\affil {$^{\rm c}$Department of Physics and Astronomy, Cardiff
University, Cardiff, 
CF2 3YB, U.K.}
\affil {$^{\rm d}$Department of Physics and Astronomy, University of Kansas, 
Lawrence, KS 66045, USA\\
Theoretical Astrophysics Center, Copenhagen, Denmark}
\affil {$^{\rm e}$ Department of Physical Sciences, Morehead State
University, Morehead, KY 40351-1689, USA}

\authoraddr{ Somnath Bharadwaj \\ Department of Physics and
Meteorology \\ IIT Kharagpur \\ 721 302 \\ India \\ e-mail:
somnath@phy.iitkgp.ernet.in} 
\begin {abstract}
We apply Shapefinders, statistical measures of ``shape'' constructed
from two dimensional partial Minkowski functionals, to study the degree of 
filamentarity in the Las Campanas Redshift Survey (LCRS). In two
dimensions, three Minkowski functionals characterise the morphology of
an object, they are: its perimeter ($L$), area ($S$), and genus. Out of 
$L$ \& $S$ a single dimensionless 
{\em Shapefinder Statistic}, ${\cal F}$, can be constructed 
($0 \leq {\cal F} \leq 1$). ${\cal F}$ acquires extreme values on a circle
(${\cal F} = 0$) and a filament (${\cal F} = 1$).
Using ${\cal F}$, we
quantify the extent of filamentarity in the LCRS by comparing our results
with a Poisson distribution with similar geometrical properties and
having the same selection function as the survey.
Our results unambiguously demonstrate that the LCRS displays a high degree
of filamentarity both in the Northern and Southern galactic sections
a result that is in general agreement with the visual appearance of the 
catalogue.
It is well known that gravitational clustering from Gaussian initial
conditions gives rise to the development of non-Gaussianity reflected
in the formation of a network-like filamentary structure on supercluster scales.
Consequently the fact that the smoothed LCRS catalogue shows properties 
consistent with
those of a Gaussian random field (Colley 1997) whereas the unsmoothed
catalogue demonstrates the presence of filamentarity lends strong
support to the conjecture that the large scale clustering of galaxies is
driven by gravitational instability.

\end {abstract}


\section{Introduction}
One of the most intriguing features of the galaxy distribution on large
scales (${\lower0.9ex\hbox{ $\buildrel > \over \sim$} } 10~h ^{-1}$Mpc.) is the
organisation of matter into geometrically complex structures often
described as being cellular, network-like,
filamentary, sheet-like, honey-comb
etc (\cite{zesh82,mel90,delgh91,sathya98}). 
The plethora of adjectives frequently used to qualify the large
scale structure of the universe underscores the difficulties inherent in
trying to quantify the morphology of the galaxy distribution. This has
as much to do with with the need for  robust statistical measures
of  the geometry and topology of large scale
structure as it does with the virtual absence, until recently, of 
redshift surveys  covering what may be regarded as a truly representative
sample of the universe. This last gap in our knowledge has been
partially filled by the Las Campanas Redshift Survey (LCRS), which
for the first time appears to  contain coherent structures whose size
is significantly smaller than the survey size. The LCRS may therefore
provide us with a statistically unbiased picture of the clustering pattern in 
the universe on very large scales.

Evidence for large scale connectivity and the presence of filaments,
pancakes and voids also comes from semi-analytic modelling of 
large scale structure and from direct N-body simulations. For instance
N-body simulations describing clustering from Gaussian initial conditions
reveal that as a distribution evolves into the non-linear regime it begins to develop non-Gaussian features characterised by the appearance, on the 
one hand of large empty regions -- voids, and on the other,
of the concentration of matter into filaments and pancakes (\cite{sc95}).
The visually prominent appearance of filaments in galaxy catalogues (CfA,
SSRS etc.) and in N-body simulations led to the endeavour to 
characterise these morphological features mathematically in order that
``real filaments'' might be distinguished from ``apparent filaments'', 
the latter arising
because the eye is prone to picking out structure when in fact there is
none. In the process it has been shown that filamentarity is a very
real feature of cosmological gravitational instability, and that the
degree of filamentarity increases as gravitational clustering advances 
(\cite{sss96,sss98b}).

Traditionally, the clustering of galaxies in groups and clusters has
been probed with some success by the two point correlation function,
$\xi(r)$, which provides an estimate of the probability (in excess of
random), of finding a galaxy at a distance $r$ from another galaxy.
A robust indicator of clustering, the two point correlation
function nevertheless reveals very little about the morphology of the galaxy
distribution, {\ie} the concentration of matter in sheets and/or
filaments and the geometrical and topological 
properties of the supercluster-void distribution as a
whole. The reason is simple, since $\xi(r)$ is the Fourier transform of
the power spectrum $P(k) = \langle\vert\delta_k\vert^2\rangle$, it does
not describe the build up of phase correlations which lead to the
emergence of non-Gaussian features in gravitating systems arising 
as a result of clustering
from Gaussian
random initial conditions.
Complete information about clustering is
formally contained in the infinite hierarchy of correlation functions
$\xi_N$, $N = 2,3,..\infty$. Although attempts have been made to
calculate $\xi_3$ and $\xi_4$, the presence of a finite number of galaxies in
any sample makes the measurement of $\xi_N$ on large scales difficult
for large $N$. Lower order correlation functions must, therefore, be
complemented  by other statistical measures sensitive to 
geometry and topology, if we wish to probe the connectedness of large
scale structure (and the associated non-Gaussianity) at a more fundamental
level.

The first statistical measures to probe the geometry of
large scale structure were percolation analysis and the genus curve
(\cite{zel82,sh83,gmd86}). 
When applied to N-body simulations and galaxy catalogues both methods pick
out departures from Gaussianity and are helpful in discerning the
presence of ``network-like''  and ``sponge-like'' features in systems
undergoing gravitational clustering. The minimal spanning tree has also
proved useful in quantifying geometrical features of large scale
structure (\cite{bbs85}).
More recently, Minkowski
functionals have added to our understanding of morphology
(\cite{mbw94,sb97}).
As demonstrated in Sahni, Sathyaprakash \& Shandarin (1998),
Shapefinders, a new shape diagnostic 
constructed out of ratios of Minkowski functionals, provides valuable 
information about the ``shape'' of a clustering pattern and can be used
with considerable advantage (in conjunction with percolation analysis)
to study issues relating to the morphology of large scale structure, such as 
the abundance of sheet-like, filament-like and ribbon-like
objects. In the present paper we shall apply Shapefinders to assess the
degree of filamentarity in the Las Camapanas Redshift Survey.

The rest of the paper is organized as follows: Sec. 2 defines the
statistics we use, Sec. 3 describes the data and the reference cataloges
we deal with, finally in Sec. 4 we present our results and conclusions.

\section{Minkowski Functionals and the Shapefinder Statistic}

In three dimensions, the four Minkowski functionals characterising the
morphology of a compact manifold are (\cite{mbw94}): 
(i) its volume, $V$, (ii) surface area, $S$, (iii) the integrated mean
curvature, $C$, and (iv) the integrated Gaussian
curvature, $G$, (equivalently the Euler characteristic or genus).
Percolation analysis 
can be accomodated within this scheme if one studies the behavior of the
volume of the largest cluster as a 
function of the filling factor for the full cluster
distribution (for mass distributions clusters being defined as 
connected objects lying above a
given density threshold).  Likewise, the genus curve can be obtained if
one plots $G$ as a function of the filling factor. 
The Shapefinder trio is constructed out of ratios of 
Minkowski functionals: $L = \frac{C}{4\pi}$, $W = \frac{S}{C}$,
$T = \frac{3V}{S}$. $L,W$ \& $T$
have dimensions of length and provide an
estimate of physical dimensions of an object such as its {\em length} $L$,
{\em width} $W$ and {\em thickness} $T$ 
(\cite{sss98a}). 
Thus $L \simeq W \simeq T$ characterises a quasispherical cluster, 
$L \gg W \simeq T$ a
filament, $L\simeq W \gg T$ a pancake and $L \gg W \gg T$ a ribbon.
Based on $L,W,T$ a pair of dimensionless
Shapefinders $\lbrace P, F\rbrace$ 
can be constructed which provide us with estimates of
the planarity $P = \frac{W - T}{ W + T}$ and filamentarity
$F = \frac{L - W}{L + W}$ of an object, $0 \leq P,F \leq 1$. 
($F = P = 0$ for a sphere, 
$F \simeq 1, P \simeq 0$ for a filament, $F \simeq 0, P \simeq 1$ for a
pancake and $F \sim P \sim 1$ for a ribbon.)

In two dimensions the partial Minkowski functionals characterizing the 
morphology of a connected region are the area $S$, 
perimeter $L$ and the number of holes in the region or genus $G$. 
The ratio $T=S/L$
characterizes the thickness of the cluster and $L$  its
extension. The ratio $L/G$ also has dimensions of length and
becomes a meaningful parameter after the onset of percolation,
characterizing
the scale of large scale structure ($L$ itself
obviously grows without limit after percolation with the growth of the survey size).

It appears that the Shapefinder statistic is a much better 
diagnostic of shape than moment-based methods, particularly when applied
to
topologically complex bodies such as isodensity surfaces occuring at
moderate density thresholds in N-body simulations and in galaxy
catalogues (\cite{sss98b}).

In two dimensions, the Shapefinder statistic simplifies to a {\em single
number} 
\beq
{\cal F} = \frac{L^2 - 4\pi S}{L^2 + 4\pi S}
\label{eq:1}
\eeq
where $L$ is the perimeter and $S$ the area of a closed two
dimensional contour. (This could be an isodensity contour in a two
dimensional galaxy
distribution, an isotherm of a hot/cold spot 
in a map of the Cosmic Microwave Background (\cite{nfsh99}), etc.) By
definition $0 \leq {\cal F} \leq 1$; ${\cal F} 
\simeq 1$ for an ideal filament (having a finite length and zero width),
and ${\cal F} \simeq 0$ for a circular disc.

The Las Campanas Redshift Survey
 contains redshifts of about 25,000 galaxies making it the largest 
magnitude limited three-dimensional catalogue of galaxies to date.
However, the six slices into which LCRS is divided are all
quasi-two-dimensional, which makes the two-dimensional shapefinder ${\cal
F}$ 
appropriate for its study.

Before applying the Shapefinder statistic to discern filamentary 
features in the LCRS,
we shall first demonstrate its effectiveness by determining ${\cal F}$
for certain eikonal shapes in two dimensions. The first object whose
shape we study 
is an ellipse of semi-major axis $a$, semi-minor axis $b$ and
eccentricity $\epsilon = \sqrt{1 - b^2/a^2}$.
The area of such an ellipse is 
\beq
S = \pi a b
\label{eq:2}
\eeq
and its perimeter is given by 
\beq
L = 4aE(\epsilon) \simeq \pi (a + b)\frac{64 - 3\lambda^4}{64 - 16\lambda^2}
\label{eq:3}
\eeq
where $E(\epsilon) = \int_0^{\pi/2}\sqrt{1 - \epsilon^2\sin^2{\theta}} d\theta$ 
is the complete elliptic integral of the second kind,
$\lambda = (a-b)/(a+b)$. The accuracy of (\ref{eq:3}) is better than $0.2\%$.
In table 1 we show the value of ${\cal F}$ as the ellipse is gradually 
deformed from an initially circular shape to a highly filamentary final
shape. As pointed out in Sahni, Sathyaprakash \& Shandarin (1998)
 other Shapefinders can also be constructed out
of the Minkowski functionals $S$ and $L$, a good example being provided by
the transformations ${\cal F}'$ = ${\cal F}^p$ and ${\cal F}'' =
\sin{(\pi{\cal F}/2)}$
which define new Shapefinders from the canonical form (\ref{eq:1}).
Transformations such as these can be made to define statistics that are
in conformity with the visual impression of morphology of archetypal
structures.

In table 1, we show results for ${\cal F}$, ${\cal F}'$ (with $p=1/2$)
and ${\cal F}''$.
>From the results presented in this Table, we see that the Shapefinder
family
${\cal F}$, ${\cal F}'$ and ${\cal F}''$ 
acquires continuous values between 0 and 1 as
a circle is deformed into a filament, and that filamentarity is more
accentuated in ${\cal F}'$ and ${\cal F}''$ than in ${\cal F}$. 
It is interesting to
contrast the behaviour of ${\cal F}$ with that of the eccentricity,
$\epsilon$, another parameter characterising shape but defined {\em
only} for an ellipse.
While the value of ${\cal F}$ grows in 
proportion to the deformation of
the ellipse (and is therefore reflective of the latter's visual shape), 
the change in $\epsilon$ is
sudden, with the result that $\epsilon$ rapidly approaches unity even 
for relatively small deformations of the ellipse.
Although the eccentricity is extremely good for measuring the small
deviations from a circle it is clearly not the kind
of behaviour one wishes to see in a ``well behaved'' shape-statistic
which should also be sensitive to large deviations from a circle.

Next, we examine a {\em topologically non-trivial} object -- 
the  region between two concentric circles with radii $R_1$ and $R_2$,
$R_2 \leq R_1$ (circular disc-with-a-hole).
In this case:
\beq
S = \pi(R_1^2 - R_2^2), ~~~~ L = 2\pi (R_1 + R_2),
\eeq
and the expression for ${\cal F}$ turns out to be very
simple
\beq
{\cal F} = \frac{R_2}{R_1}.
\eeq
As the radius of the hole $R_2$ shrinks to zero, the object becomes a
circular disc with ${\cal F} \simeq 0$, in the other extreme case when
$R_1 \sim R_2$ the disc-with-a-hole reduces to a circular
filament having ${\cal F} \simeq 1$
(rather like the mythological serpent eating its tail). 
One should note that the presence of circular symmetry would lead a
moment-based statistic to wrongly
declare such an object as being circular or homogeneous (\cite{sss98b}).

Results of increasing the hole size by increasing $R_2$ are shown
in Table 2 for ${\cal F}$, ${\cal F}'$ and ${\cal F}''$. They are in broad
agreement with those obtained earlier for the ellipse.

We therefore find that the two dimensional 
Shapefinder statistic
 gives sensible results when applied to 
both simple and topologically 
complicated eikonal shapes. This conclusion is supported by  
results obtained for three dimensional Shapefinders by Sahni et al.
(1998) and Sathyaprakash, Sahni \& Shandarin (1998).
In this paper we shall apply ${\cal F}$
to study filamentarity in the Las Camapanas Redshift Survey (the results of
applying ${\cal F}'$ and ${\cal F}''$ 
are qualitatively similar and will not be discussed separately).

\section{Analysis.}
\subsection{The Las Camapanas Redshift Survey.}

The Las Camapanas Redshift Survey (LCRS) contains approximately 25,000
galaxies with known redshifts. The survey region is divided into six
slices, three each in the Northern (N$_1$, N$_2$, N$_3$)
 and Southern (S$_1$, S$_2$, S$_3$) Galactic hemispheres.
The slices are strips of the sky $1.5^\circ$ thick and 
$80^\circ$ wide which are separated
by $3^\circ$ and probe a distance of up to
$600~h^{-1}$Mpc. (h is the value of the Hubble parameter in units of 
100 km~sec$^{-1}$Mpc$^{-1}$). 
The centers of the three northern slices are at declinations $-3^\circ$
(N$_1$),
$-6^\circ$ (N$_2$) and $-12^\circ$ (N$_3$), whereas the three southern
slices are 
centered at $-39^\circ$ (S$_1$), $-42^\circ$ (S$_2$) and $-45^\circ$
(S$_3$). The survey is 
complete to limiting magnitude $m = 17.75$ (for more details of the LCRS
survey see \cite{shectman96}).

In order to minimise selection and projection effects 
we apply the Shapefinder statistic to a volume limited subsample of the
LCRS derived from the dense central region $200 \leq R \leq
400~h^{-1}$ Mpc.  
(The LCRS selection
function peaks at $R \simeq 200 h^{-1}$Mpc. and the region $200 \leq R
\leq 400~h^{-1}$ Mpc corresponds to the densest part of the survey.)
In order to ascertain the statistical significance of our results we
compare them with a statistically homogeneous two-dimensional Poisson 
distribution with 
identical selection function and geometry as the survey 
and corrected for projection
effects (for more details, see \cite{sy98}).

Each slice is embedded in a $560 \times 260$ grid with resolution
$1~h^{-1} {\rm Mpc}$.  This enables us to define galaxy locations on a two
dimensional lattice. Lattice cells containing a galaxy are said to be
occupied or {\em filled}, and  lattice  cells with no galaxies are
referred to as {\em empty}. The cells which lie outside the boundaries
of the survey are eliminated from the lattice.  

 Next, we proceed to ``grow'' structure (or coarse-grain) by means of
the following iterative procedure: At each step filled cells are made to
grow isotropically (\ie in all directions) by a single unit of the mesh
size. Thus any cell neighboring a filled cell (one containing a
galaxy) in any of eight possible directions is also designated ``filled''.
This algorithm leads to the growth of 
filled cells at each successive iteration. 

The procedure of sequentially coarse-graining a LCRS slice is shown in
Figure 1 for the N1 slice. Incidentally, it also demonstrates what
happens when  
the size of the dots in the wedge diagram plots are arbitrarily chosen. 
Selecting the size of the dots or the linking length one can
emphasize ``desired'' features of the distribution such as filamentarity,
connectivity, the size of the greatest supercluster or something else. 
So far no reason has been suggested in the literature for a particular 
choice of the linking 
length (except one corresponding to percolation transition) and therefore 
we use the linking length or the actual size of the dots for the
purpose of parameterization. 
 
The associated {\em filling
factor} (FF), defined as the fraction of filled cells in the total slice,
increases from a small initial value to almost unity when all the
cells in the LCRS slice have inter-connected into one big
all-pervading cluster. At any given value of FF ({\em i.e.,} at any given
iterative step) we determine clusters using a 
friends-of-friends algorithm: any filled cell sharing a common side with
another filled cell is its friend. All such ``friends'' define a cluster.
The shape of individual clusters defined in this manner can be
analysed using the Shapefinder statistic at different values of the
filling factor. Large scale connectedness of galaxies in
LCRS is also revealed by studying the Largest Cluster Statistics (LCS)
characterizing the percolation transition 
as a function of the filling factor
(LCS is defined as the ratio of the area of the largest cluster to the
total
area of all clusters, see \cite{sy98}).

It is also necessary to point out that measurements of 
structure in galaxy catalogues  can in principle depend upon the 
observational strategy, particularly on the proximity
of fibres mounted on a CCD. Ones intrinsic inability to pack fibres
sufficiently close together in order to measure redshifts of all galaxies in
a dense field might lead to the undersampling of galaxies in clusters. 
However since filaments are much lower density objects than clusters 
their detection will not be as 
sensitive to the spacing between fibres with the result that
most of the results presented in the 
present paper are expected to be sufficiently robust. 
(For the rare case of a filament 
aligned along the line of sight and therefore viewed ``head on'' 
this effect could lead to undercounting of galaxies along the core
of the filament and therefore 
weaken the signal for filamentarity. Our present estimates of filamentarity
in LCRS should therefore be treated as a lower bound to the extent of
filamentarity in the ``real universe''.)

\subsection{Shapefinders on a grid}

The Shapefinders 
(\ref{eq:1})
were defined for continuous contours. For contours
on a grid this definition must be modified.
It is easy to see that the lattice version of (\ref{eq:1}) is
\beq
{\cal F}_1 = \frac{L^2 - 16 S}{L^2 + 16 S}
\label{eq:4}
\eeq
(Note that the factor 
$4\pi$ in (\ref{eq:1}) has been replaced by 16 in (\ref{eq:4}).)
Consider a rectangle defined on a grid with 
sides $n\times l$ and $m\times l$, respectively, where $l$ is the
inter-grid spacing. 
It is easy to see that in this case ${\cal F}_1$ reduces to
\beq
{\cal F}_1 = \frac{(n - m)^2}{(n + m)^2 + 4nm}
\label{eq:5}
\eeq
as a result ${\cal F}_1 = 0$ for $n = m$ (a square), and 
${\cal F}_1 \rightarrow 1$ for $n \gg m$ (a filament).

It is interesting that one can define a second Shapefinder
statistic which is also 
suitable for determining shapes of contour lines defined on a grid.
Again, if $l$ is the inter-grid spacing we have
\beq
{\cal F}_2 = \frac{L^2 - 16S}{(L - 4l)^2}
\label{eq:6}
\eeq
so that, for the rectangle $(n,m)$ 
\beq
{\cal F}_2 = \frac{(n - m)^2}{(n + m - 2)^2}.
\label{eq:7}
\eeq
We see that once more ${\cal F}_2 = 0$ for $n = m$ (a square), and 
${\cal F}_2 \rightarrow 1$ for $n \gg m$ (a filament). A specific
feature of ${\cal F}_2$ is that a rectangular contour of unit width 
$(n,1)$ is always
delared to be a filament regardless of its length $n$, since, 
substituting $m = 1$ in (\ref{eq:7}) we get
${\cal F}_2 = (n-1)^2/(n-1)^2 = 1$.

One might question the need for introducing two separate statistics to 
study shapes on a lattice. 
The reason for this is related to the following question:
Consider an ``ideal'' one-dimensional filament having a finite
length but zero breadth. How would one go about representing this object
on a grid? Depending upon how one answers this question one arrives
at either ${\cal F}_1$ or ${\cal F}_2$. One might argue
that the discrete analogue of an ideal filament would be a rectangular
object with width $\sim l$, (\ie width =  mesh size) 
in which case ${\cal F}_2$ gives the correct continuum
limit: ${\cal F}_2 = {\cal F} = 1$. 
On the other hand one might equally argue that a filament having
width $< l$ is impossible to correctly define on a grid for which the
Nyquist wavelength determines an effective small scale ``resolution cutoff''. 
Therefore, if one
is confined to a grid one must necessarily differentiate between
objects having 
identical widths ($\sim l)$ but varying in length, in this
case the correct statistic to use is ${\cal F}_1$.

In this paper, we report on work carried out using both ${\cal F}_1$ and
${\cal F}_2$. As we shall show, ${\cal F}_1$ and ${\cal F}_2$
contain complementary information about filamentarity, as a result both
prove to be very useful shape-diagnostics for large scale structure.

\cite{nfsh99} used yet another approach to measuring
the Minkowski functionals on a grid. Studying the 4-year COBE maps
they assumed 
a smooth underlying field and, therefore, used a linear interpolation
scheme  
for measuring the perimeters and areas of the clusters.  

\section{Results and Conclusions.}

A general statistical quantity that can be constructed out of the
Shapefinder ${\cal F}$ is the functional $n({\cal F}|A,FF)$ which
describes the number density of disjoint regions of area $A$ 
having given values of the
filamentarity ${\cal F}$ and located above a given
density threshold characterised by the
filling factor $FF$. However a good simple indicator of filamentarity
is provided by the averaged quantity 

\beq
\langle {\cal F}^{(d)}(FF)\rangle
= \frac{\int\int n({\cal F}|A,FF){\cal F} A^d~dA~ d{\cal F}}
{ \int\int n({\cal F}|A,FF) A^d~dA~ d{\cal F}}
\label{eq:7a}
\eeq

 describing particular moments
of filamentarity in the distribution.
(Each cluster of area $A$ is
 identified using a nearest neighbors algorithm at a given
value of the coarse-graining.)

Clearly the extent of filamentarity in a survey is strongly influenced
by the morphology of its largest and 
most massive members: a supercluster contributes
more to the overall texture of large scale structure than an individual galaxy.
We therefore  examine moments with $d > 1$ thereby giving greater weight to
larger objects. Since our results 
are qualitatively similar for $d = 2,3,4...$
we show only results for $d = 2$ and refer to the statistic
$\langle {\cal F}^{(2)}(FF)\rangle$ simply as $F$.
For the discrete sample under consideration (\ref{eq:7a}) reduces to
\beq
F \equiv F_{1,2} = \frac{\sum_i A_i^2{\cal F}_{1,2}^i}{\sum_i A_i^2}, ~~~~~~
0 \leq F_{1,2} \leq 1
\label{eq:8}
\eeq
where the sum is evaluated at a given value of the filling factor. 
$A_i$ is the area of a given cluster and ${\cal F}_{1,2}^i$ is the filamentarity of the
$i^{th}$ cluster obtained by using (\ref{eq:4}) for ${\cal F}_1$ or (\ref{eq:6}) for 
${\cal F}_2$.

In Figures 2 and 3, we show the extent of filamentarity as a function of the 
filling factor for three northern and three southern slices of LCRS
(the volume limited slice contain from 800 to 1,600 galaxies).
To assess the statistical
significance of our results we compare $F$
with the corresponding
quantity derived from four independent
 Poisson randomizations of the given LCRS slice. From Figures 2 and 3
 it is clear
that for moderate values of the filling factor the extent of
filamentarity in {\em all} six slices of LCRS is {\em significantly greater}
than in the Poisson sample. This is true for $F$ constructed from 
both ${\cal F}_1$ and
${\cal F}_2$. Comparing Fig. 2 (obtained using ${\cal F}_1$) with Fig.
3
(obtained using ${\cal F}_2$) we find that the difference between LCRS
and the Poisson sample shows up at much lower filling factors for
${\cal F}_2$. 
The reason for this is simple, as we demonstrated earlier, ${\cal F}_2$ 
is designed to be 
more successful in picking out smaller filaments than ${\cal F}_1$.
At low FF, small filaments appear to be much more abundant in LCRS
than they are 
in the Poisson sample, as a result ${\cal F}_2$ easily discerns
filamentarity in LCRS at small FF. Together ${\cal F}_1$ and ${\cal
F}_2$ probe the extent of filamentarity over a wide range of FF.

Complementary information emphasising 
the ``connectivity'' of a distribution is revealed
by percolation analysis (\cite{sy98}).
At small FF each slice contains a large number of
distinct clusters
whose shape must be determined individually. As the
FF increases, neighboring clusters begin to merge leading to a decrease
in the total number of clusters in the sample and to the emergence of a
single dominant ``supercluster''. With increased FF we find that
the largest cluster shows a very rapid increase in size as a result of
which it soon spans the entire survey region. This corresponds to the
onset of percolation. We determine ${\cal F}$ for individual clusters
and the area of the largest cluster at each level of coarse graining.
As $FF \rightarrow 1$ the percolating supercluster fills the entire
slice,
the value of ${\cal F}$ then drops to a small value which describes the
shape of the survey region. (If the slice were an exact square the value
${\cal F}$ would approach 0 as $FF \rightarrow 1$.)
In Fig. 4 we plot the Largest Cluster Statistic 
(LCS is the fractional size
of the largest cluster relative to all clusters at a given value of
coarse graining) against 
FF (the fraction of filled cells in the total area of the slice).
Comparing these results with those of the Poisson distribution we clearly
see that the growth in the largest cluster is more rapid in LCRS 
than it is in the randomized
catalogue (constructed with the same selection function and number of
galaxies and therefore having identical geometrical properties to LCRS) 
confirming the earlier results of Shandarin and Yess (1998). 

To summarise we have shown
that clustering of galaxies in LCRS is significantly non-Gaussian: 
it is dominated by filaments and its global geometry is network-like 
(in the parlance of \cite{sy98}).
Our analysis also indicates that
Shapefinders recognize non-Gaussian features in LCRS
at lower filling factors than the percolation curve and that 
Shapefinders and percolation analysis supplement each other in 
discerning the geometrical properties of the galaxy distribution on
large scales.

Finally we note that at $FF \sim 1$ the extent of filamentarity in LCRS
declines (from a value close to unity) 
and becomes smaller than in the Poisson catalogue. 
The decrease in $F$ in both LCRS and the Poisson catalogue is simply
due to the fact that all clusters in the sample have now merged into 
a single ``super-duper'' cluster which resembles a
quasi-homogeneous object with several holes (empty regions). 
Such an
object when probed using the Shapefinder statistic appears very
filamentary both in LCRS and the Poisson sample. However
as $FF \rightarrow 1$ the holes gradually get filled and 
$F$ drops to a small value which describes the shape of the
survey region. 
We do not attach much 
significance to results shown in Fig.2 and Fig.3 at $FF > 0.6$
because at such large values of the filling factor 
the percolating supercluster spans the 
entire sample region. As a result at large $FF$ 
some geometrical properties of the supercluster begin
to be progressively determined by the size and geometry of the sample 
volume and cease to be reflective of the (formally infinite) 
galaxy distribution from which 
the finite sample is drawn.

The largest cluster in two dimensions can be characterized by two
parameters: the thickness of the cluster wall separating nearby voids $T=S/L$ and the length of the
cluster boundary per void $L/|G|$.
The cluster perimeter $L$ itself is not a good measure
because for the percolating cluster it diverges as the area of the survey grows
without bound. On the other hand the parameter $L/|G|$ defined for
the percolating cluster proves to be remarkably stable
over a wide range of linking lengths.
Fig. 5 shows the length of the cluster boundary per void ($L/|G|$) as a function of
wall thickness ($S/L$) for both the LCRS and reference catalogues.
The stability of the statistic $L/|G|$ is reflected by the fact that
while the thickness of the supercluster grows by a factor of three from
$\approx 5~h^{-1}$Mpc to $\approx 16~h^{-1}$Mpc  the mean value of $L/|G|$ 
decreases only
marginally:
from $\approx 110~h^{-1}$Mpc to $\approx 70~h^{-1}$
Mpc in the LCRS catalogue.
The structure in the reference Poisson catalogues is of considerably
smaller size  than in the LCRS. This is reflective of the existence of large scale structure
in the galaxy distribution marked by a percolating network of filaments separating large voids.

Although Fig. 5 gives the scale of the structure, we wish to stress that
it is not the diameter of the empty cells seen in Fig. 1. The parameter 
$L/|G|$ has a meaning of ``length of the supercluster contour line per
void''. In our case 
the mean number of voids is large: about 30 in the LCRS and about 130 in the
randomized catalogues (at wall thickness of $\approx 10~h^{-1}$Mpc). This allows us to say
that the sample is sufficiently large and the average diameter of the voids
(at thickness $\approx 10~h^{-1}$Mpc) is about $80/\pi \approx 25~h^{-1}$Mpc.
(Large values of $|G|$ for a given cluster indicate a large number of holes
and consequently a greater porosity of the cluster. This applies 
particularly to the percolating supercluster for which $G$ can be
very large and negative.)

It is interesting to note that an analysis of LCRS using the genus curve
has shown that at large values of the smoothing radius, LCRS
reveals statistical properties which are consistent with those of a Gaussian
random field (\cite{colley97}). Our results on the other hand (and those
of \cite{sy98}) show that the {\em unsmoothed} LCRS catalogue displays
strongly non-Gaussian features. In this context, it is worthwhile to
reiterate that N-body gravitating systems clustering from Gaussian
initial conditions rapidly develop non-Gaussian features reflected both
in a network-like structure and
by a growth of filamentarity in the morphology of clusters and
superclusters (\cite{sss96,sss97,sss98b}). The effects of smoothing such a
distribution on large scales would be to add greater weight to
 the clustering of
long range modes still in the linear regime, and hence statistically
distributed in the manner of a Gaussian random field (\cite{ds92,sc95}).
When viewed in this context the results of this paper
(and those of \cite{sy98}), 
together with
the results of Colley (1997), 
appear to provide
strong support for the scenario in which the large scale structure of
the universe evolved via gravitational clustering from Gaussian initial
conditions predicted (for instance) by Inflationary Models of the very
early universe.

Acknowledgements. S. Shandarin acknowledges the support of NSF-EPSCoR 
grant, GRF grant at the University of Kansas and from TAC Copenhagen.
We wish to thank an anonymous referee for comments that led to an improvement in the
manuscript.

\clearpage

\begin {figure}[ht]
\begin{center}	
\plotone{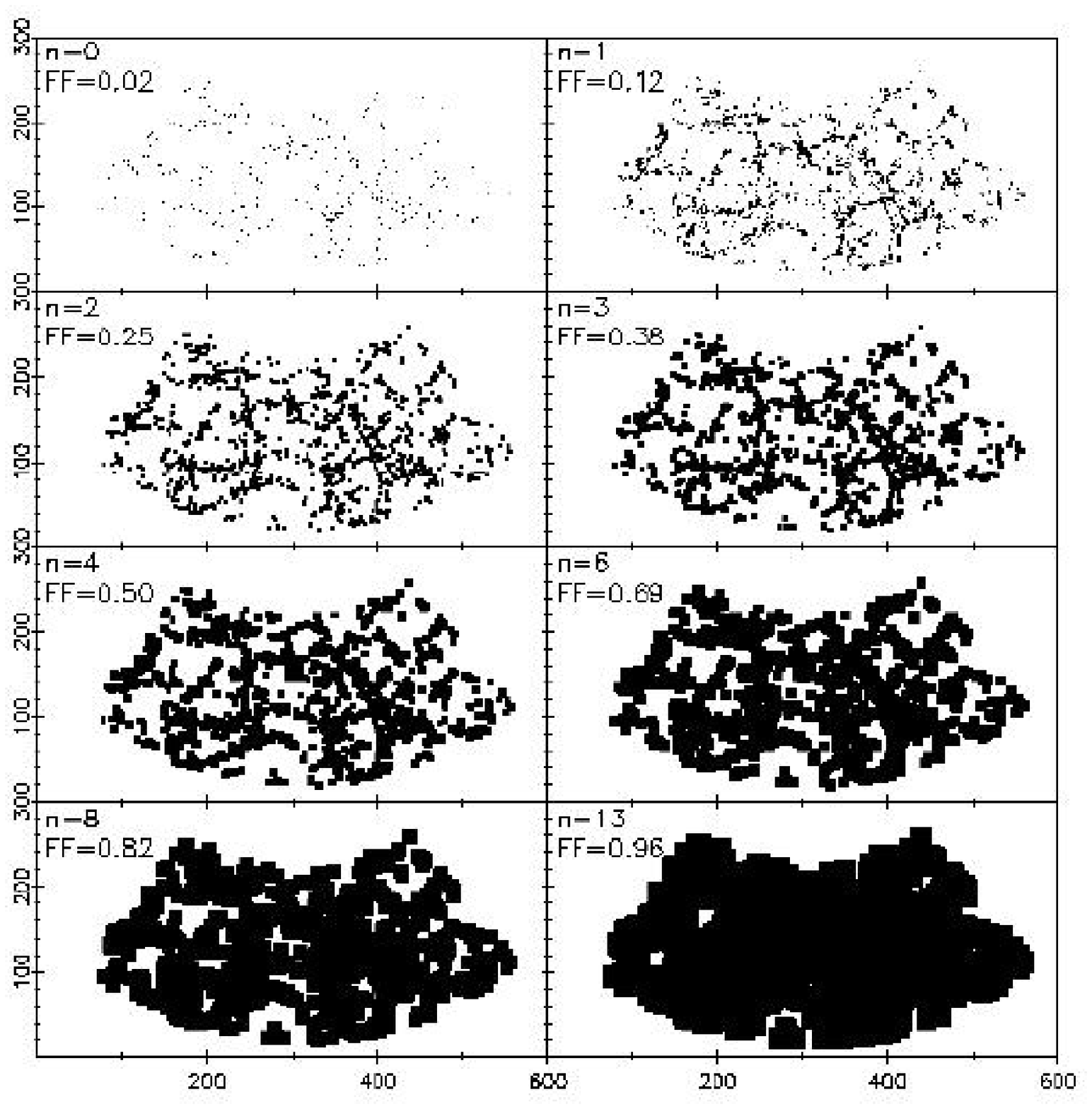}

\caption {\small The first Northern LCRS slice shown at different values of 
coarse-graining (n). The value of the filling factor (FF) at each
level of coarse-graining is also shown. Here ${\rm n} = 0$ shows
galaxies in the 
original $N_1$ slice without coarse graining. The axis units are
$h^{-1}$ Mpc.}

\end{center}
\end {figure}

\begin {figure}[ht]
\begin{center}
\plotfiddle{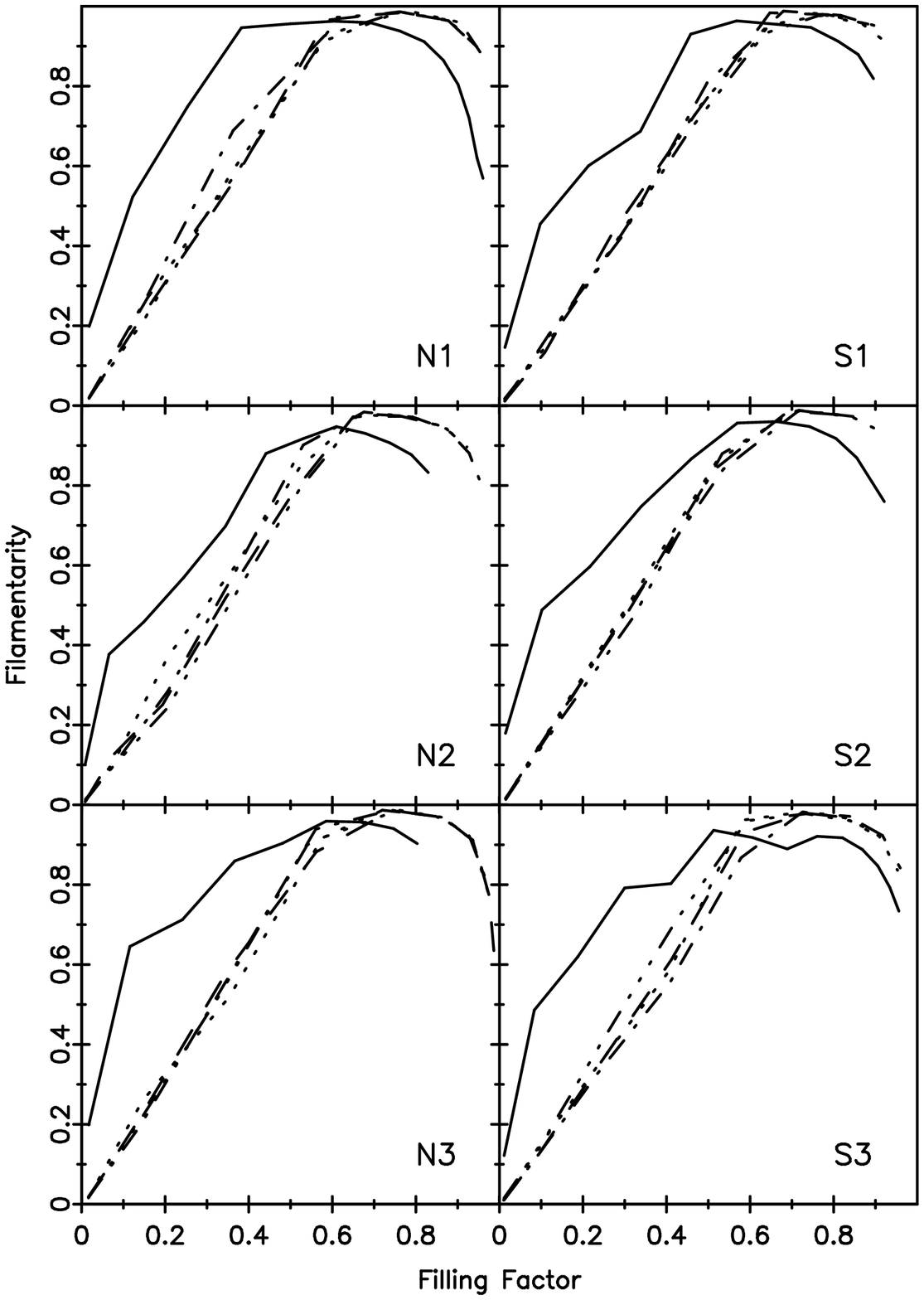}{6in}{0}{70}{70}{-180}{-30}
\caption {\small The filamentarity statistic $F$ is shown plotted
against the filling factor FF for each slice in the  LCRS  (solid
line).  For comparison we also show $F$ for four
independent random Poisson fields for each of the slices (dashed and dotted
lines). The shape-statistic $F$ 
shown here is constructed from the discrete Shapefinder ${\cal F}_1$.}

\end{center}
\end {figure}

\begin {figure}[ht]
\begin{center}
\plotfiddle{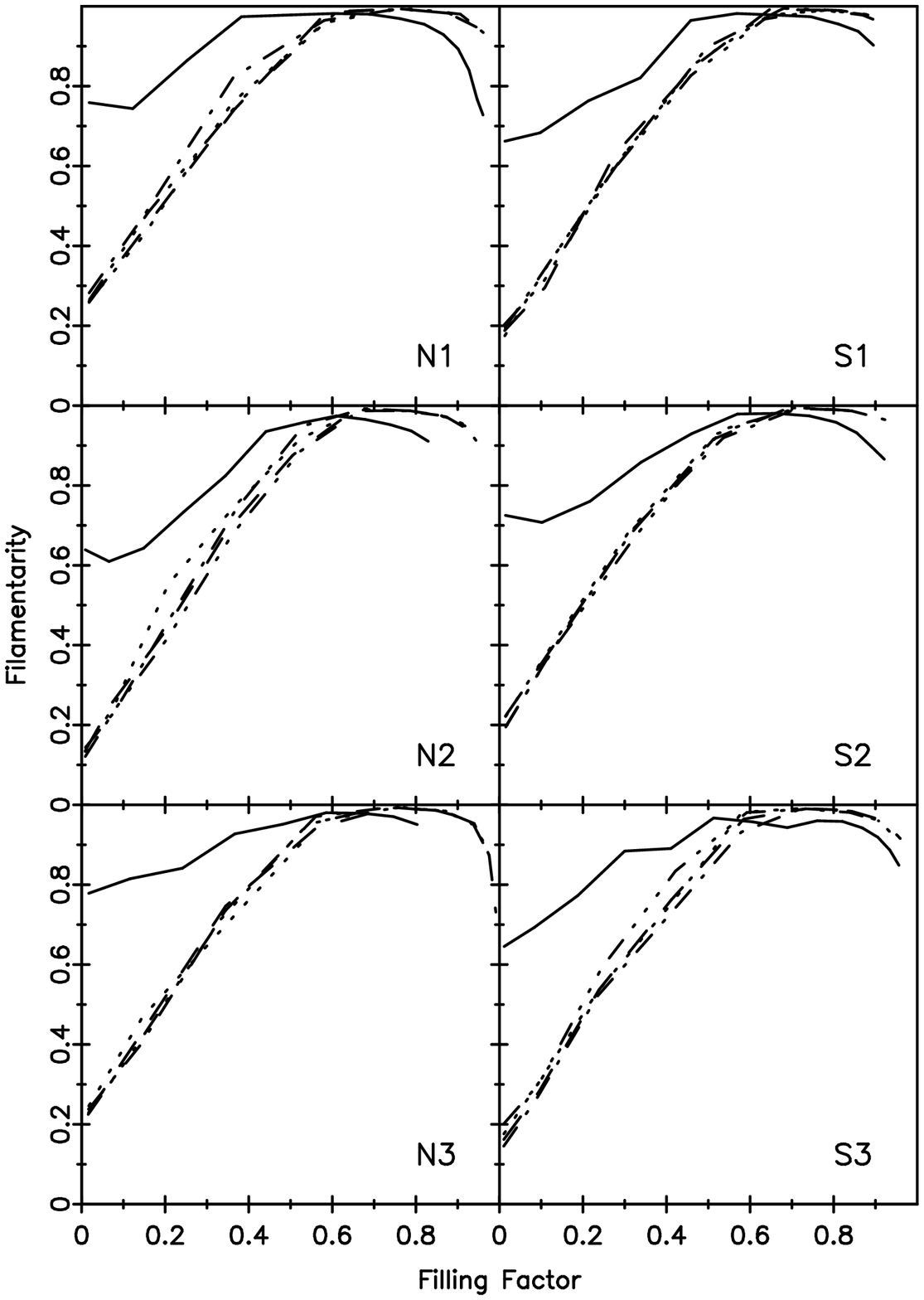}{6in}{0}{70}{70}{-180}{-30}
\caption {\small The filamentarity statistic $F$ is shown plotted
against the filling factor FF for each slice in the  LCRS  (solid
line).  For comparison we also show $F$ for four
independent random  Poisson fields for each of the slices (dashed,dotted
lines). The shape-statistic $F$ 
shown here is constructed from the discrete Shapefinder ${\cal F}_2$.}

\end{center}
\end {figure}

\begin {figure}[ht]
\begin{center}
\plotfiddle{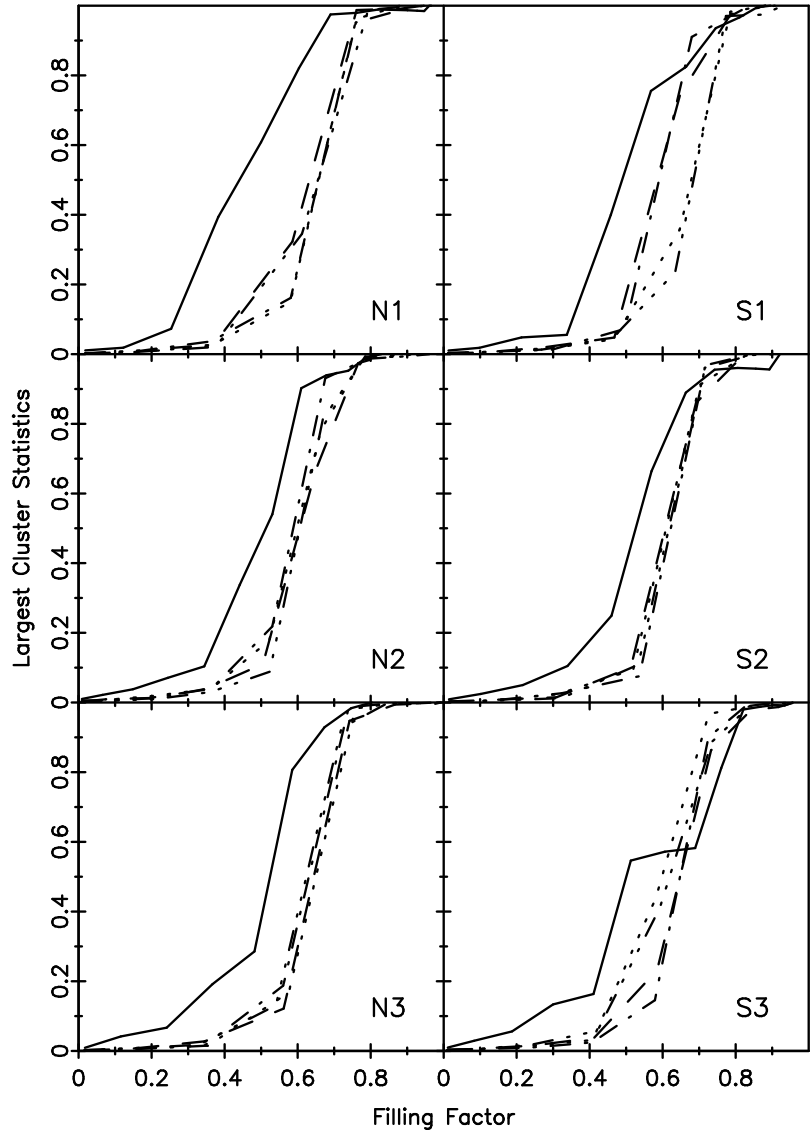}{6in}{0}{70}{70}{-180}{-30}
\caption {\small The largest cluster statistics (LCS) is shown plotted
against the filling factor FF for each slice in the  LCRS  (solid
line).  For comparison we also show $(LCS)$ for four
independent random Poisson fields for each of the slices (dashed,dotted
lines). }
\end{center}
\end {figure}

\begin {figure}[ht]
\begin{center}
\plotfiddle{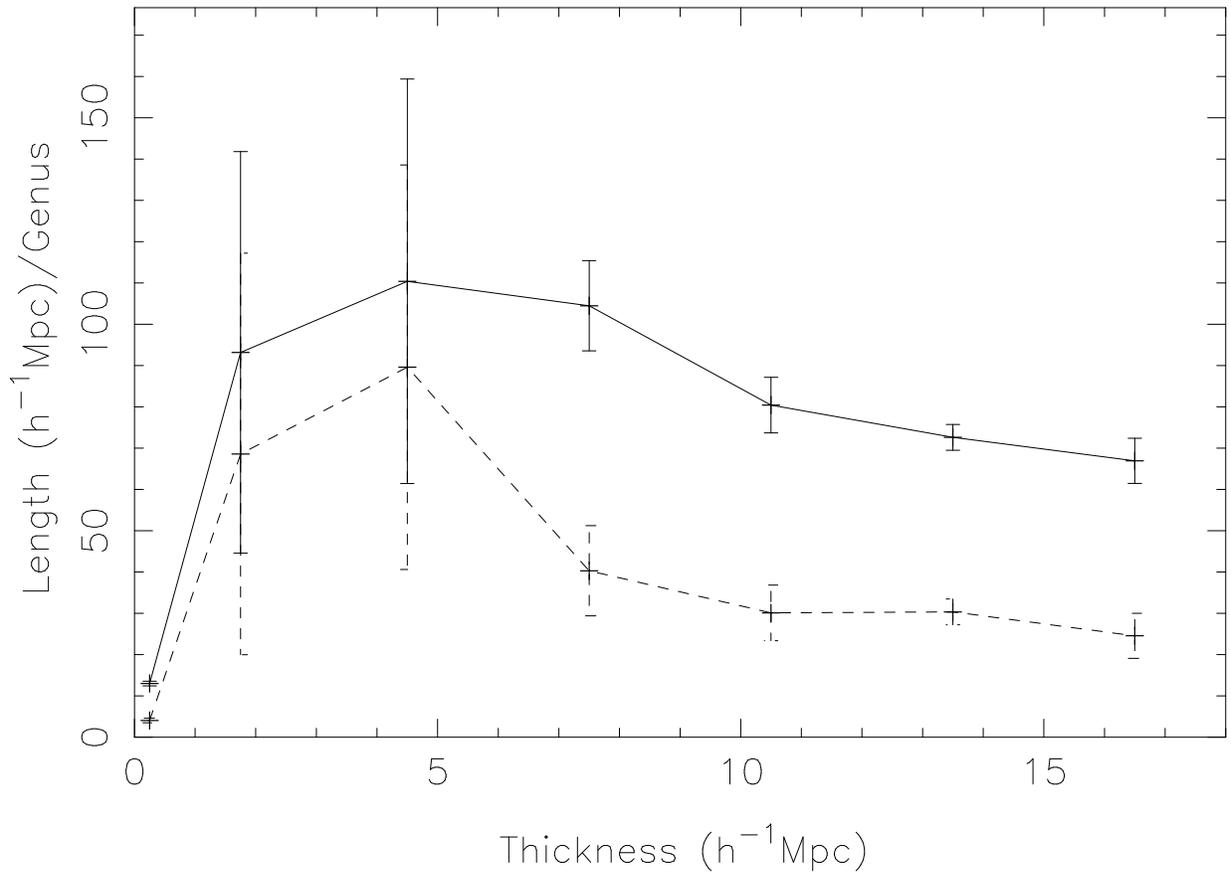}{6in}{-90}{70}{70}{-250}{450}
\caption {\small The solid curve  shows the ratio length/genus  of the
largest cluster   
plotted  as a function of  the thickness of the largest cluster. The
values of the thickness from the different slices have been divided
into bins of $ 3  h^{-1}$ Mpc 
and the average value of the length/genus in each bin  has been
plotted as the 
solid curve. The error-bars show $1-\sigma$ fluctuations. For
comparison we have shown the same quantity for the random Poisson
fields using dashed lines. 
}

\end{center}
\end {figure}

\clearpage

\vspace{2in}
\begin{table}
\begin{center}
\caption{Shapefinders ${\cal F}$, ${\cal F}'$ \& ${\cal F}''$
describe the shape of
 an ellipse having semi-major axis $a$, 
semi-minor axis
$b$ and eccentricity $\epsilon$.}
\bigskip
\begin{tabular}{llllll}
\tablerule
$a/b$ & $\epsilon$ & ${\cal F}$ & ${\cal F}' = {\cal F}^{1/2}$ & ${\cal
F}'' = \sin{(\pi {\cal F}/2)}$\\
\tablerule
$1$ & 0 & 0 & 0 & 0\\
$3$ & 0.94 & 0.20 & 0.45 & 0.31\\
$5$& 0.98 & 0.38 & 0.62 & 0.56\\
$10$& 0.99 & 0.61 & 0.78 & 0.82\\

\tablerule
\end{tabular}
\label{table:shape}
\end{center}
\end{table}

\begin{table}
\begin{center}
\caption{Shapefinders ${\cal F}$, ${\cal F}'$ \& ${\cal F}''$
describe the shape of the
region between two concentric
circles with radii $R_1$ \& $R_2$ (circular disc-with-a-hole).}
\bigskip
\begin{tabular}{lllll}
\tablerule
$R_1, R_2$ & ${\cal F}$ & ${\cal F}' = {\cal F}^{1/2}$ & ${\cal F}'' =
\sin{(\pi 
{\cal F}/2)}$\\
\tablerule
$10, 0$ & 0 & 0& 0\\
$10, 2$ & 0.20 & 0.45 & 0.31\\
$10, 4$ & 0.40 & 0.63 & 0.59\\
$10, 6$ & 0.60 & 0.77 & 0.81\\
$10, 8$ & 0.80 & 0.89 & 0.95\\
$10, 9$ & 0.90 & 0.95 & 0.99\\
\tablerule
\end{tabular}
\label{table:shape1}
\end{center}
\end{table}

\end {document}